\documentclass[sigplan]{acmart}
\setcopyright{none}
\renewcommand\footnotetextcopyrightpermission[1]{} 
\AtBeginDocument{%
  }

\usepackage{amssymb}
\usepackage{subfig}
\usepackage{wrapfig}
\usepackage{float}
\usepackage{multirow, graphicx}
\usepackage{algorithm}
\usepackage{algpseudocode} 
\usepackage{amsmath}
\usepackage{xspace}

\usepackage{url}

\usepackage{breakurl}

\usepackage{sty/widetext} 

\usepackage{stackrel}

\ifdefined\TTSTYLETARGETreport
\usepackage[
    backend=bibtex,
    maxbibnames=100,
    firstinits=true,
    bibstyle=numeric,
    citestyle=numeric
]{biblatex}
\fi

\usepackage{pifont}
\newcommand{\cmark}{\ding{51}}%
\newcommand{\xmark}{\ding{55}}%

\usepackage{todonotes}

\newcommand{\ignore}[1]{}
\usepackage{graphicx, multirow}

\usepackage{tabularx, multirow, rotating} 
\usepackage{subfig} 
\usepackage{algorithm} 
\usepackage{algpseudocode} 
\usepackage{ulem} 
\usepackage{color} 

\usepackage{listings} 
\lstdefinestyle{Oracle}{basicstyle=\ttfamily,
                        keywordstyle=\lstuppercase,
                        emphstyle=\itshape,
                        showstringspaces=true,
                        }
\makeatletter
\newcommand{\lstuppercase}{\uppercase\expandafter{\expandafter\lst@token
                           \expandafter{\the\lst@token}}}
\newcommand{\lstlowercase}{\lowercase\expandafter{\expandafter\lst@token
                           \expandafter{\the\lst@token}}}
\makeatother

\usepackage{tikz}

\newif\ifboldnumber

\algrenewcommand\alglinenumber[1]{%
  \footnotesize\ifboldnumber\bfseries\fi\global\boldnumberfalse#1:}

\usepackage{color, colortbl}
\definecolor{Gray}{gray}{0.9}
\definecolor{LightCyan}{rgb}{0.88,1,1}
\usepackage[first=0,last=9]{lcg}

\usepackage{colortbl}
\usetikzlibrary{calc}
\usepackage{zref-savepos}

\newcounter{NoTableEntry}
\renewcommand*{\theNoTableEntry}{NTE-\the\value{NoTableEntry}}

\usepackage{makecell}

\begin{document}

\title{Enabling Cost-Effective Blockchain Applications via Workload-Adaptive Transaction Execution}
\subtitle{A Case Study on Saving Fees for Write-intensive Accounts}


\author{Yibo Wang}
\affiliation{%
 \institution{Syracuse University}
 \city{Syracuse}
 \state{New York}
 \country{USA}
}
\email{ywang349@syr.edu}

\author{Yuzhe Tang}
\affiliation{%
  \institution{Syracuse University}
  \city{Syracuse}
  \state{New York}
  \country{USA}
}
\email{ytang100@syr.edu}

\begin{abstract}
  As transaction fees skyrocket today, blockchains become increasingly expensive, hurting their adoption in broader applications.

This work tackles the saving of transaction fees for economic blockchain applications. The key insight is that other than the existing ``default'' mode to execute application logic fully on-chain, i.e., in smart contracts, and in fine granularity, i.e., user request per transaction, there are alternative execution modes with advantages in cost-effectiveness.

On Ethereum, we propose a holistic middleware platform supporting flexible and secure transaction executions, including off-chain states and batching of user requests. Furthermore, we propose control-plane schemes to adapt the execution mode to the current workload for optimal runtime cost.

We present a case study on the institutional accounts (e.g., coinbase.com) intensively sending Ether on Ethereum blockchains. By collecting real-life transactions, we construct workload benchmarks and show that our work saves $18\%\sim{}47\%$ per invocation than the default baseline while introducing $1.81\sim{}16.59$ blocks delay.

\ignore{
Ethereum is the second largest blockchain platform and charges a very high fee for transaction execution. This work presents OCIA, a blockchain optimization scheme for write-intensive accounts, to optimize the cost of transferring Ether. OCIA stores the state of an account off-chain and dynamically decides when to upload the off-chain states to the on-chain storage. Leveraging batch and off-chain states, OCIA saves $18\%\sim{}47\%$ per invocation than baseline while having $1.81\sim{}16.59$ blocks delay.
}

\end{abstract}



\maketitle
\pagestyle{plain}
\newcommand{\yz}[1]{\footnote{\textcolor{red}{(Yuzhe: #1)}}}

\providecommand{\eoa}{\textsf{EOAlgo}\xspace}

\definecolor{mygreen}{rgb}{0,0.6,0}
\definecolor{mymauve}{rgb}{0.88,0.69,0}
\lstset{ %
  backgroundcolor=\color{white},   
  basicstyle=\scriptsize\ttfamily,        
  breakatwhitespace=false,         
  breaklines=true,                 
  captionpos=b,                    
  commentstyle=\color{mygreen},    
  deletekeywords={...},            
  escapeinside={\%*"}{"*)},          
  extendedchars=true,              
  keepspaces=true,                 
  keywordstyle=\color{blue},       
  language=Java,                 
  morekeywords={*,...},            
  numbers=left,                    
  numbersep=5pt,                   
  numberstyle=\scriptsize\color{black}, 
  rulecolor=\color{black},         
  showspaces=false,                
  showstringspaces=false,          
  showtabs=false,                  
  stepnumber=1,                    
  stringstyle=\color{mymauve},     
  tabsize=2,                       
  title=\lstname,                  
  moredelim=[is][\bf]{*}{*},
}

\section{Introduction}

Ethereum~\cite{me:eth2} is a blockchain platform that supports smart contracts. Ethereum charges a very high fee for transaction execution. One cause of the high fee is that data storage and processing are replicated to all nodes across a large network. Another cause is due to the basic principle of economy -- raising user demand in transactions under the limited supply of block space (i.e., causing more users to bid for a single block slot). As a result, the Ether price has increased $27$ times in the last two years (as of Mar. 2022). The high fees have real-world consequence: Ethereum clients are scared away and are forced to switch to other blockchains.

There are several cost saving schemes on blockchain that have been studied in the existing literature. iBatch~\cite{DBLP:conf/sigsoft/WangZLTCLC21} supports batching multiple smart-contract invocations into one transaction so that the base fee is amortized. GRuB~\cite{DBLP:conf/middleware/LiTCYXX20} stores data off-chain in data-feed workloads and dynamically replicates the data on the blockchain. Layer-two protocols~\cite{me:lightning,DBLP:journals/corr/MillerBKM17,DBLP:journals/corr/abs-1804-05141,DBLP:conf/ccs/DziembowskiFH18} are built on top of the blockchain as extensions. They process the application logic off-chain to increase throughput and reduce gas fee. iBatch supports the invocations between any kinds of accounts, while is limited as it doesn't support Ether transferring. GRuB is designed only for data-feed workloads and doesn't support Ether transfer as well. Layer-two protocols support Ether transferring with the payment network. It saves Gas cost only for the transactions between institutional accounts. 

While iBatch is under smart contract workloads and GRuB is under data feed workloads, it remains an open research problem: Can one optimize the cost of institutional client transferring Ether by leveraging batch and off-chain states?

We tackle this problem by presenting OCIA, a blockchain \underline{O}ptimization s\underline{C}heme for \underline{I}nstitutional \underline{A}ccounts via selectively placing states off the blockchain. OCIA is a control-plane scheme that maps the upper-layer transaction workloads (Ether transfers) to the underlying data plane that extends blockchains with batch transactions (i.e., iBatch) and off-chain states (i.e., GRuB). This work extends the workload to new, more common workloads, that is, Ether transfers from institutional accounts to average accounts. 

We propose an (offline) dynamic optimization algorithm that decides when to upload the off-chain states to the on-chain storage (i.e., to convert an off-chain account to on-chain one) in order to maximize the batch size and minimize the cost.

We discover the institutional accounts who send the most transactions in one day Ethereum transaction history. We then analyzed the transaction history related with $3$ institutional accounts as case study workloads including Coinbase, Ethermine, Crypto.com. The dynamic optimization algorithm makes one transaction contain more Ether transferring to amortize the cost while introducing a delay. We analyze the cost under the workload of write-intensive accounts with the dynamic optimization algorithm. The result shows OCIA saves $18\%\sim{}47\%$ per invocation than baseline while has $1.81\sim{}16.59$ blocks delay.

\begin{table*}[!htbp] 
  \caption{Cost Optimization Schemes}
  \label{tab:optimization_schemes}
  \centering{\small
  \begin{tabularx}{0.8\textwidth}{ |l|X|l|l|l| }
    \hline 
    Scheme & \multicolumn{2}{l|}{Workload} & Approach & Supported pattern \\ \cline{2-3}
    & Supported & Ether transfer &  &  \\ \hline
  iBatch & Smart contract & \xmark & Batched requests & \makecell[l]{Any-any}   \\ \hline
  GRuB & Smart contract (Data Feed)  & \xmark & Off-chain states & N/A  \\ \hline
  Layer-2 & Ether transfer & \cmark & Payment networks & \makecell[l]{Institutional - \\institutional}  \\ \hline
  OCIA & Ether transfer & \cmark & Batch + Off-chain storage & Institutional - avg \\ \hline
\end{tabularx}
}
\end{table*}

\section{Preliminary: Ethereum transaction model}

In Ethereum, there are two types of blockchain accounts, externally owned account (or EOA) and smart contract account (CA). An account is associated with certain on-chain “states”. In the case of EOA, the state is the account balance. In the case of CA, the state additionally includes the storage variables in the smart contract. 

Each transaction contains multiple instructions and each instruction has a specific cost that is counted in Gas. The transaction sender pays the fee based on the amount of Gas to compensate for the computational effort of the Ethereum network. Each transaction has a base fee that is $21000$ Gas covering the cost of verifying the signature of the sender as well as uploading the transaction data.

\section{Execution Modes and Cost}
\begin{figure}[!ht]
  \centering
  \includegraphics[width=0.495\textwidth]{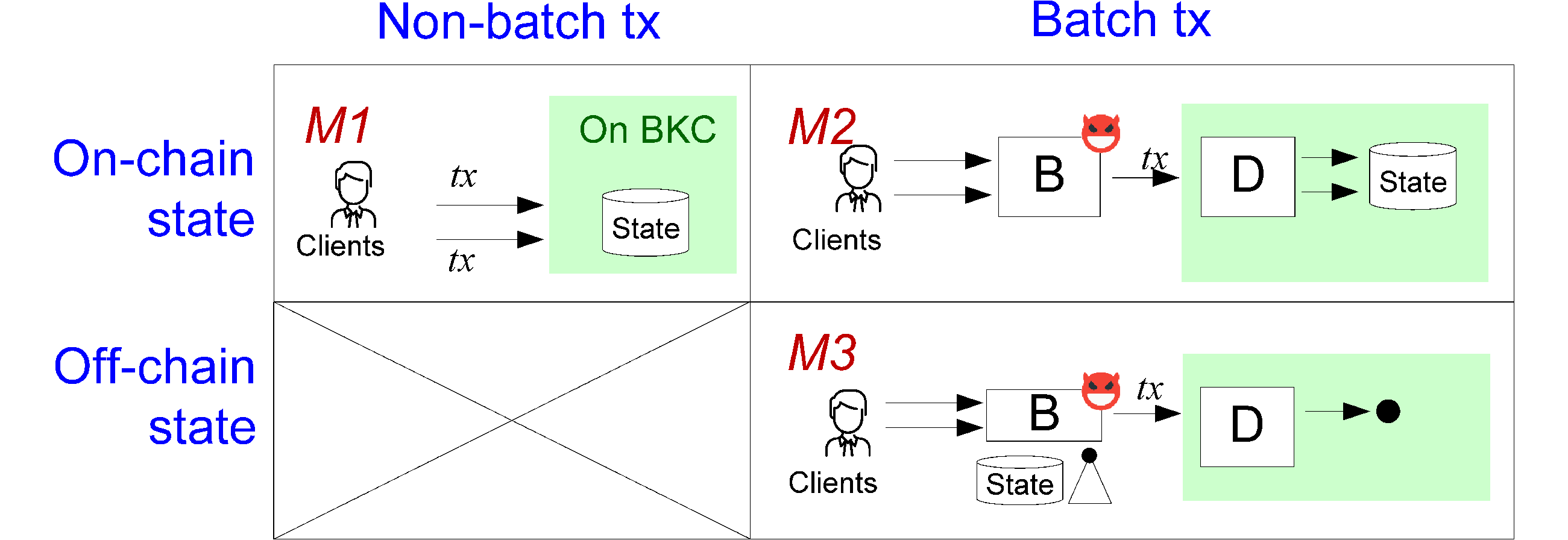}
  \caption{Transaction execution modes.}
  \label{fig:execution_modes}
  \end{figure}

Blockchain transactions can be executed in different equivalent modes. By default, account states are on-chain and each client request is executed in a dedicated transaction. Alternatively, states can be placed off-chain, and multiple requests can be batched in one transaction. Figure~\ref{fig:execution_modes} shows different execution modes in M1, M2, and M3. Note that we discard the conceptual model dealing with non-batch transaction and on-chain state, because this model does not save costs in our analysis.

In the following, we present M1, M2, and M3 in more detail. 

\subsection{Non-batched Tx and On-chain States ($M1$)}

All accounts' states are stored on the blockchain (including EOA's balance and CA's program data). An EOA's request to transfer Ether is executed in a dedicated transaction.

To establish the cost model, we consider an over-simplified workload named $W1'$. In $W1'$, an institutional account A0 sends a sequence of $N$ Ether-transferring transactions in $L$ blocks. Under $W1'$, the cost under the mode M1 is $21000 * N$, where $21000$ is the cost of sending an Ether-transferring transaction on Geth v1.10.17/Solidity 0.7.0. Thus, the per-transfer cost is $C_1 = 21000$
\subsection{Batched Tx and On-chain States ($M2$)}

In Execution Mode $M2$, accounts' states are stored on the blockchain. The EOA's request is batched with other requests, submitted at similar time, in a large transaction. As shown in Figure~\ref{fig:execution_modes}, the batch transaction is sent to a smart-contract, called Dispatcher, that dispatches the individual EOAs' requests to their corresponding on-chain states.
Specifically, the
iBatch scheme~\cite{DBLP:conf/sigsoft/WangZLTCLC21} materializes $M2$ with an untrusted Batcher off-chain that is desgined with the resilience against request-replay attacks.

Under the workload $W1'(N, L)$, we use $NW$ to denote the amount of Ether transfer of $Tx1$ in L blocks shown in Figure~\ref{fig:system_workload}.
The cost with $M2$ is $21000 * L + NW * [68 * (20+32) + 7500]$, where $68 * ( 20 + 32 )$ is the transaction cost storing data field\footnote{$68$ is Gas cost per byte in the data field and the $20$ and $32$ respectively refer to the number of bytes storing the receiver address and value in an Ether transfer.} and $7500$ is the cost of calling an internal Ether transfer. Thus, the per-transfer cost is $C_2 = 21000 L/NW + 11036$.

\subsection{Batched Tx and Off-chain States ($M3$)}
In Execution Mode $M3$, accounts' states are stored off the blockchain and clients' requests are batched in transactions.

\noindent{\bf Sender/receiver off-chain}: The system is built on top of iBatch by placing the account states to an untrusted off-chain service. When a request, say Alice transferring Ether to Bob, is batched in a transaction, the on-chain smart contract needs to authenticate Alice's and Bob's old balances and update them, and compute and update the new digest including Alice's and Bob's updated balances. Thus, the transaction's data includes the Merkle proofs authenticating Alice's and Bob's old balances against the old digest.

\noindent{\bf Sender on-chain, receiver off-chain}: The $M3$ places the receivers' states to an off-chain server and the state changes in a Hash log by expanding the Merkle tree instead of updating the state in off-chain directly. When receiving requests, say $A_0$ transferring Ether to $A_1$ and $A_2$, we add $dA1$ and $dA2$ as new leaf nodes and keep the original balance unchanged. In each block, we upload the new digest and the state changes on-chain. The on-chain smart contract leverages the updating information and the old digest to authenticate and update the new digest. Under workload $W1'$, the cost is $21000 * L + NW * [68 * (20 + 36) + 222] + (68 * 32 + 5000) * L$ where $222$ is the cost of SHA3 Keccak-256 operation of two $32$ bytes digest, $68 * 32$ is the cost of uploading the new digest in the call data field and $5000$ is the cost of updating the digest for each block in the on-chain smart contract. Thus, the per-transfer cost is $C_3 = 28176L/NW + 4030$.

\begin{figure}
  \centering
  \includegraphics[width=0.425\textwidth]{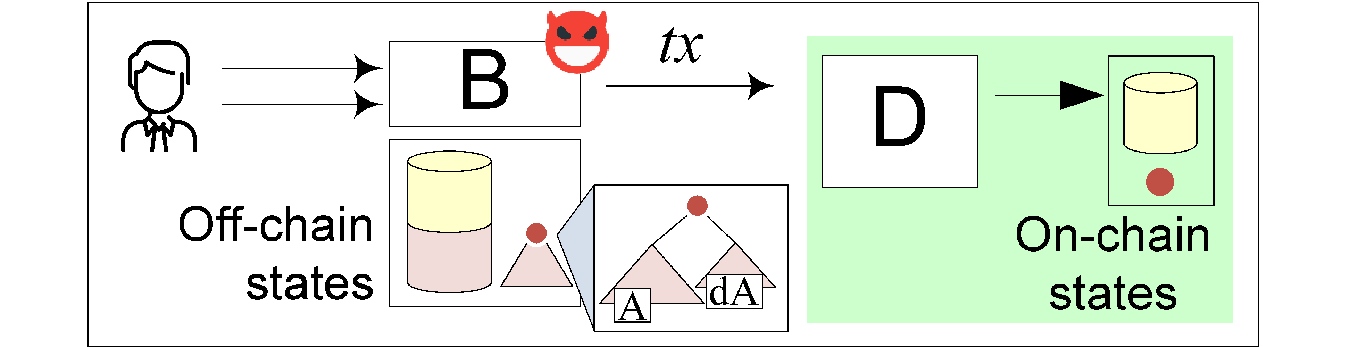}
  \caption{Dynamic execution mode for ``sender on-chain receiver off-chain'' workloads.}
  \label{fig:system_m4}
\end{figure}
\section{Cost Analysis for Workload $W1$}

\begin{figure}
  \centering
  \includegraphics[width=0.245\textwidth]{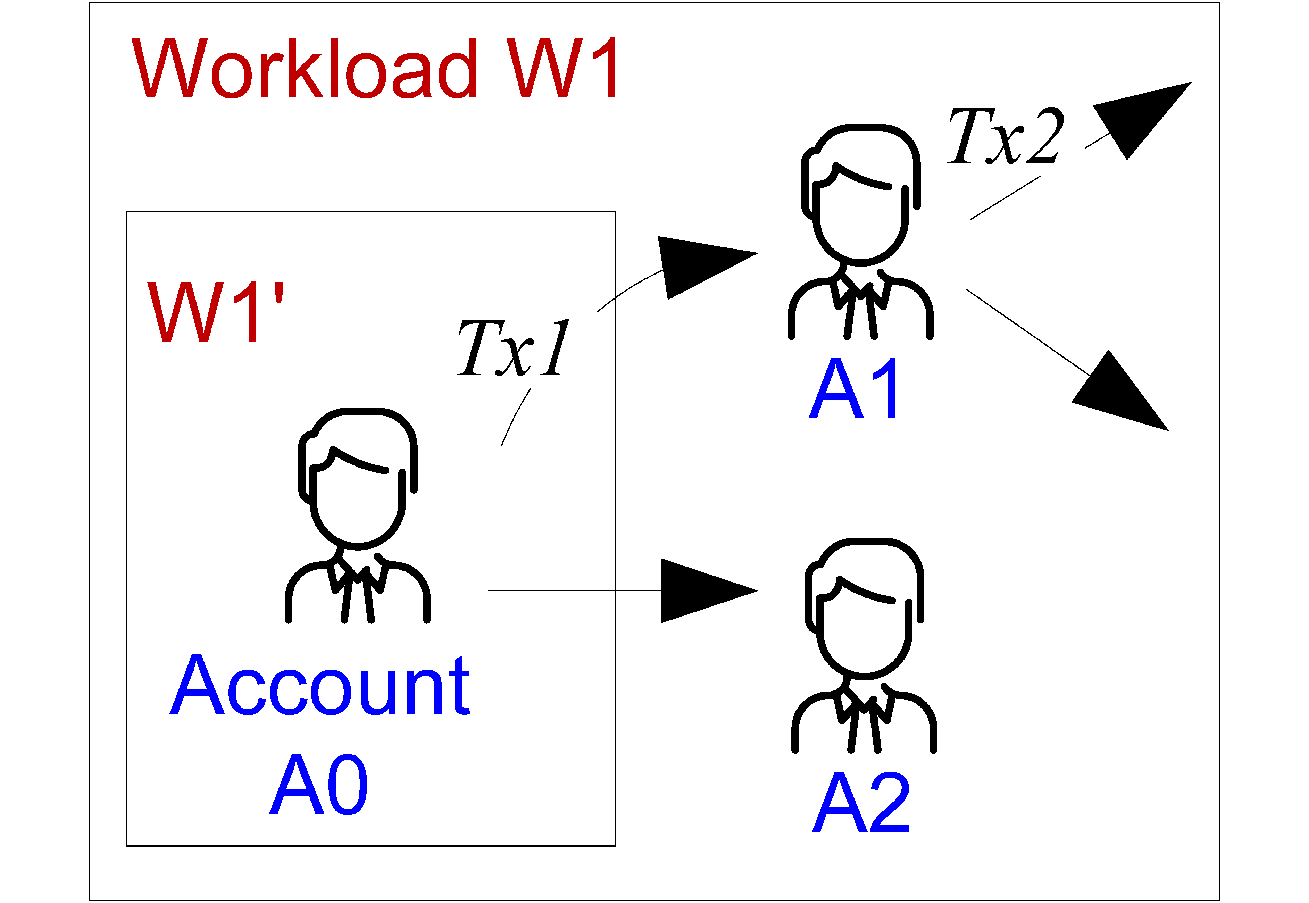}
  \caption{Target workload $W1$ to optimize.}
  \label{fig:system_workload}
\end{figure}

The target workload pattern we optimize is shown in Figure~\ref{fig:system_workload}. Given an institutional account $A_0$, there are three transactions of interest in our workload: the transactions sent by $A_0$, the transactions received by $A_i$, and the transactions sent by $A_i$.

{\bf $M1$'s costs under workload W1}: For $M1$, the on-chain states of $A_i$ are updated in $Tx1$. Hence, the cost of $Tx2$ is the same as a normal Ether transferring. Thus, the cost of $W1$ is $C_1 = 21000 * (NW + NR)$.
Here, $NW$ is the amount of Ether directly transferred from institutional account $A_0$, and $NR$ is the amount of Ether transferred within two hops from the insttitutional account (in Figure~\ref{fig:system_workload}).

{\bf M2's costs under workload $W1$}: For M2, the on-chain states of $A_i$ for $\forall{i\leq{1}}$ are updated in Tx1. Hence, the cost of Tx2 is the same as a normal Ether transferring transaction, which is $21000$. Thus, the cost of $W1$ is $C_2 = 21000*L + 11036 * NW + 21000 * NR$

{\bf $M3$'s costs}: When $A_i$ for $\forall{i\leq{1}}$ under $M3$ sends Ether to another account, the off-chain server push all the $A_i$ related states on-chain, includes the balance state and the updating value in every blocks. The off-chain server also uploads the Merkle proofs of each related account's off-chain state. The on-chain smart contract leverages the Merkle proofs to authenticate the updated balance of $A_i$.  Under workload $W1'$, say $A_i$ with an initial balance in off-chain server, it receives $V_i$ Ether in $L$ blocks with $NW$ transactions and it sends Ether to other accounts with $NR$ transactions. The cost is $21000L + NR * [68 * (20 + 36 + 65) + 6600 + 7500] + NR * 68 * (20 + L) * 32 * L + NR * (20 + L) * 222 + (68 * 32 + 5000) * L$,
where $68 * (20 + 36 + 65)$ is the cost of uploading the receiver address, transfer value and sender's signature, $6600$ is the cost of the signature verification and $7500$ is the cost of an internal Ether transfer. $68 *(20+L)* 32*L$ is the cost of uploading the 20+L-level Merkle proof for L blocks. $(20+L)*222$ is the cost of 20+L SHA3 Keccak-256 operation to verify the digest. Thus, the cost of $W1$ is
$C_3 = 28176L + 4030 * NW + 22328 * NR + NR * 68 * (20 + L) * 32 * L +NR * (20 + L) * 222$.

\section{Cost Evaluation}

\subsection{Discover Institutional Account}

We download the Ether transferring transaction history and find the top senders who send most Ether transferring transactions as the institutional accounts on March 1st, 2022. We list the top $5$ senders in Table~\ref{tab:institutional-senders}. We then analyze Coinbase, Ethermine and Crypto.com as study cases. The workloads contain the transactions, say Tx1, that are sent from an institutional sender, say A0, to its receivers, say $A_i$. The workloads also contain the transactions of Ether transfer that are sent from $A_i$, say Tx2, after they receive the Ether from A0.

\begin{table}[!htbp]
  \caption{Top-5 institutional senders}
  \label{tab:institutional-senders}
  \centering{\small
  \begin{tabularx}{0.475\textwidth}{ |l|X|X|X| }
    \hline 
  Sender & Amount percentage ($* 100\%$)& Value (Ether)  & Value percentage ($* 100\%$)\\ \hline
  Ethermine (ea674f) & $4.81$ & $5306$ & $0.24$  \\ \hline
  Coinbase 3 (ddfabc) & $2.08$ & $18816$ & $0.87$   \\ \hline
  Coinbase 4 (3cd751) & $2.08$ & $19127$ & $0.88$   \\ \hline
  Coinbase 5 (b5d85c) & $2.05$ & $24248$ & $1.13$  \\ \hline
  Coinbase 6 (eb2629) & $2.05$ & $20667$ & $0.96$   \\ \hline
  
  \end{tabularx}}
\end{table} 

\subsection{Characterizing Collected Workloads}

{\bf Coinbase:} Coinbase is a cryptocurrency exchange platform. We treat $6$ Coinbase Ethereum accounts as one institutional account. We analyze the workload and count the number of $NW$ and $NR$. The average amount of $NW$ and $NR$ per block are $12.30$ and $25.62$.

Under $W1$, the amortized cost of $M1$ is $21000$. The amortized cost of $M2$ is $18321$, which saves $12.75\%$ than baseline. The amortized cost of $M3$ is $51159$, which costs $143.62\%$ more than baseline.

{\bf Ethermine:} Ethermine is a big miner in Ethereum. The average amount of $NW$ and $NR$ per block are $4.65$ and $4.88$. The amortized cost of $M2$ is $18341$, which saves $12.66\%$ than baseline. The amortized cost of $M3$ is $42143$, which costs $100.68\%$ more than baseline.

{\bf Crypto.com:} Crypto.com is also a cryptocurrency exchange platform. The average amount per block of $NW$ and $NR$ are $1.49$ and $10.89$. The amortized cost of $M2$ and $M3$ are $21497$ and $66698$, which cost $2.37\%$ and $217.61\%$ more than baseline.
\begin{figure}[!ht]
  \centering
  \includegraphics[width=0.475\textwidth]{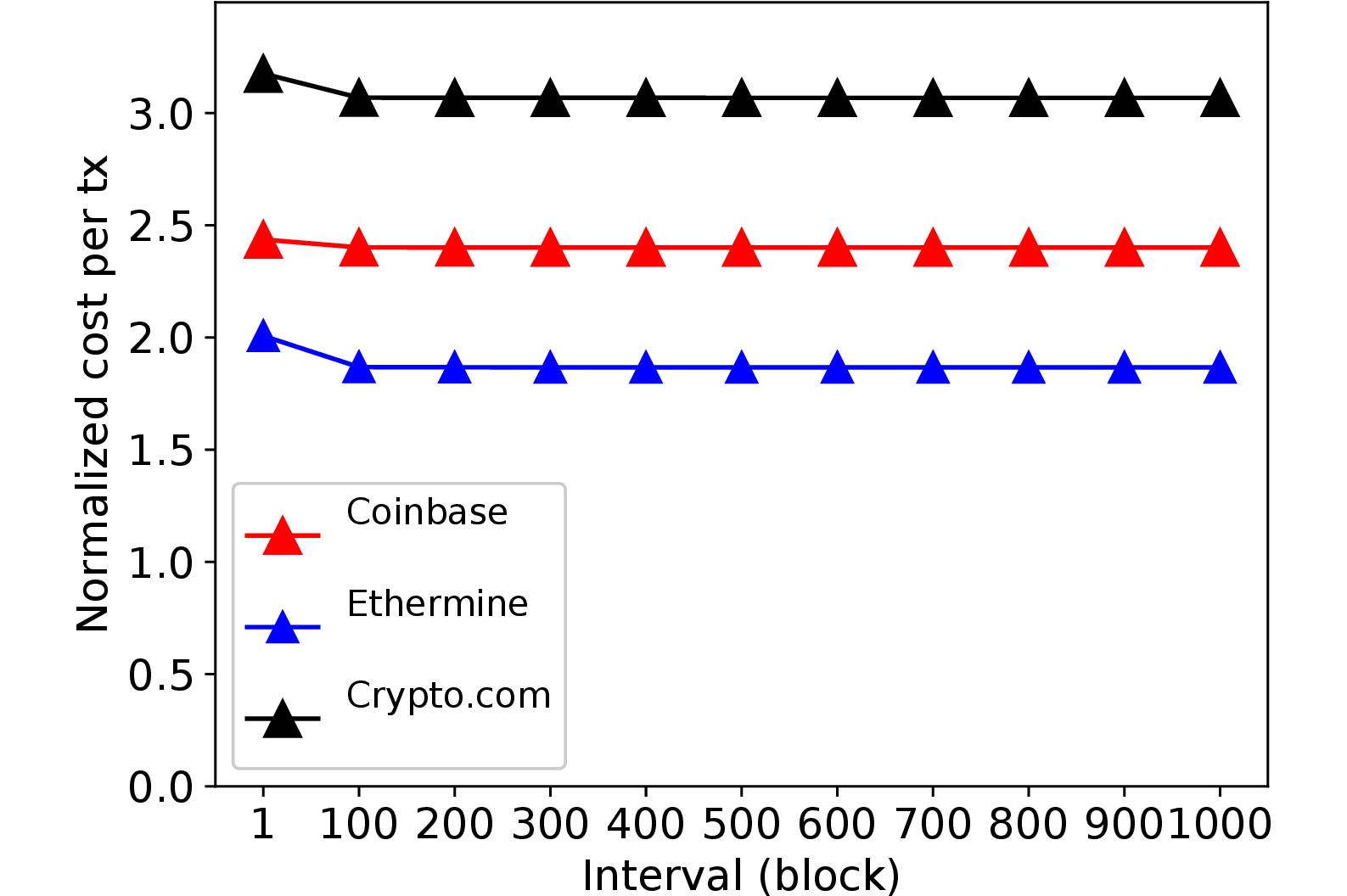}
  \caption{Amortized Cost with blocks delay.}
  \label{fig:workload-delay}
\end{figure}
\subsection{Cost Evaluation Under Real Workloads With Delay}

While the above delay-free method costs more than baseline, we introduce delay to update the digest and execute the Ether transferring. For each $K$ blocks, we upload a new digest and the updated state on-chain, specifically the receiver's addresses and the transfer values. For every Ether transferring under $W1$ workloads, it takes $K - 1$ blocks delay at most. 

We analyze the workloads of Coinbase, Ethermine and Crypto.com with various block intervals. The figure~\ref{fig:workload-delay} illustrates the analyzed result of the workloads with $100$ to $1000$ blocks intervals. The cost per transaction has a small decrease as the delay increases, while it is still larger than the baseline. With $1000$ block intervals, the cost of OCIA is $86.6\%\sim{}206.75\%$ more than baseline.

\subsection{Cost Evaluation Under Write-intensive Workloads}
The amortized cost per transaction of $M3$ depends on the number of $NW$ and $NR$. We find the accounts that meet the conditions $C_3 - 21000 < 0$ as write-intensive accounts, that is the accounts whose $NW$ and $NR$ subject to $NR < (8485 * NW - 14088) / 25843$ and $NW > 1.66$. We get the transaction history of the write-intensive accounts as write-intensive workloads. $13.00\%$ of accounts in Ethermine workloads are write-intensive accounts. The workloads of Coinbase and Crypto.com have $2.14\%$ and $0.56\%$ write-intensive accounts.

We dynamically upload the off-chain states to the on-chain storage when the $NW$ and $NR$ in the off-chain state meet the write-intensive condition. With this policy, the cost of each transaction is less than baseline while it introduces delay. We analyzed the average cost of each transaction and the average delay of the write-intensive workloads in table~\ref{tab:cost-delay-write-intensive-worloads}.

\begin{figure}[!htp]
  \subfloat[ Normalized cost with various policies ]{
    \includegraphics[width=0.25\textwidth]{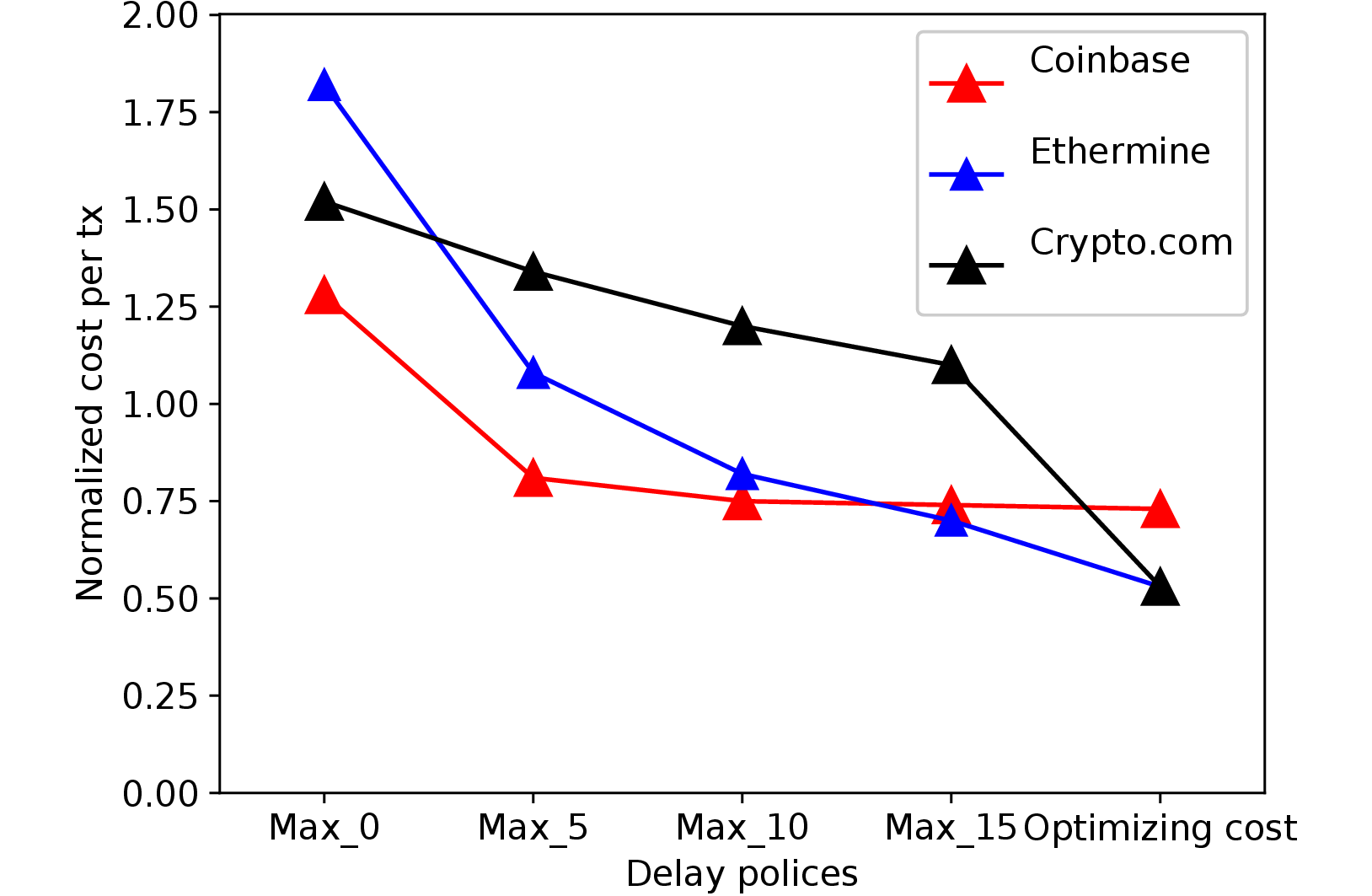}
    \label{fig:cost-delay-tradeoff-cost}
  }
  \subfloat[ Average delay with various policies ]{
    \includegraphics[width=0.25\textwidth]{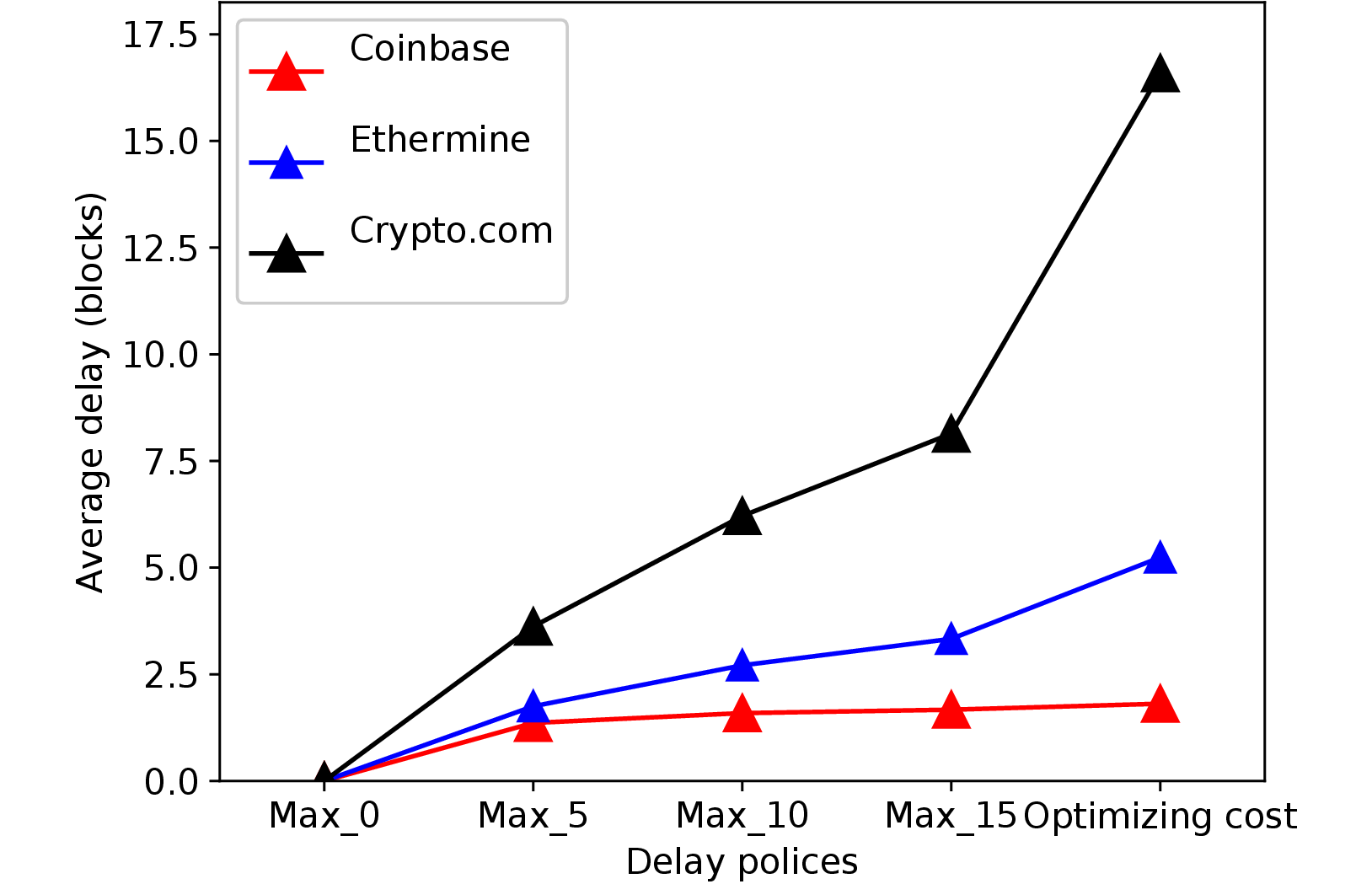}
    \label{fig:cost-delay-tradeoff-delay}
  }
  \caption{ Trade-off between cost saving and delay.}%
  \label{fig:cost-delay-tradeoff}
\end{figure}

The above {\bf optimizing cost policy} saves cost while the delay is relatively large. We propose various delay policies to make trade-off between cost saving and delay. {\bf Max\_0 policy}: updates the on-chain state with no delay. It updates the on-chain state in the same block interval with the Ether transferring requests. There is no delay introduced in Max\_0 policy, while it hardly achieves a positive cost saving because few requests are batched in the transaction. {\bf Max\_5 policy}: updates the on-chain state with at most $5$ blocks delay. The state of Ether transferring requests are first stored off-chain. If the $NW$ and $NR$ meet the conditions of saving cost within $5$ blocks, we send a transaction to update the on-chain state. Otherwize, we update the on-chain state no matter what the $NW$ and $NR$ is to make sure the delay is not larger than $5$. {\bf Max\_10 policy}: updates the on-chain state with at most $10$ blocks delay. {\bf Max\_15 policy}: updates the on-chain state with at most $15$ blocks delay.
The trade-off of cost saving and delay with various delay policies is illustrated in figure~\ref{fig:cost-delay-tradeoff}.

\begin{table}
  \caption{Cost and delay of write-intensive worloads}
  \label{tab:cost-delay-write-intensive-worloads}
  \centering{\small
  \begin{tabularx}{0.47\textwidth}{ |X|l|l|l| }
    \hline 
   & Coinbase & Ethermine  & Crypto.com\\ \hline
   Normalized cost & $0.73$ & $0.53$ & $0.82$  \\ \hline
   Average delay (block) & $1.81$ & $5.24$ & $16.59$   \\ \hline 
  \end{tabularx}}
  \end{table}

\section{Related Work}
{\bf Batching smart contract invocations}: iBatch~\cite{DBLP:conf/sigsoft/WangZLTCLC21} focuses on enabling secure batching of smart-contract invocations to optimize the base fee of a Ethereum transaction. iBatch uses a smart contract to dispatch each invocation, it doesn't support Ether transfer between two Ethereum accounts and it requires the application smart contract rewrote to support iBatch. In iBatch protocol, all the states are stored on-chain, while OCIA stores the state off-chain to save the storage write cost. 

{\bf Storing data off-chain}: GRuB~\cite{DBLP:conf/middleware/LiTCYXX20} optimizes the cost in feeding external data to a blockchain. GRuB stores data off-chain in write-intensive workloads and dynamically replicates the data on the blockchain. The saving of this dynamic replication is workload-specific in data feeding, while OCIA supports more general Ether transferring workloads. 

{\bf Layer-two}: Layer-two protocols~\cite{me:lightning,DBLP:journals/corr/MillerBKM17,DBLP:journals/corr/abs-1804-05141,DBLP:conf/ccs/DziembowskiFH18} is built on top of the blockchain. Layer-two protocols support Ether transferring with the payment network. Payment network~\cite{me:lightning,DBLP:journals/corr/MillerBKM17,DBLP:journals/corr/abs-1804-05141} remains only two transactions (e.g., opening and closing a channel) on-chain to increase throughput and reduce cost. After opening a channel, it executes the application logic and stores the payment states off-chain. It updates the states on-chain when the channel is closed. 

\section{Conclusion}

This paper presents OCIA, a scheme that stores the write-intensive account states off-chain to save the transaction cost. OCIA applies a trustless scheme to upload the states on-chain when it is read by on-chain transactions. The result shows the OCIA saves $18\%\sim{}47\%$ under write-intensive workloads. 

\section{Acknowledgement}
The authors are partially supported by the National Science Foundation under 
Grants CNS2139801, CNS1815814 and DGE2104532.

\bibliographystyle{ACM-Reference-Format}
\bibliography{bkc}


\begin{thebibliography}{7}


\ifx \showCODEN    \undefined \def \showCODEN     #1{\unskip}     \fi
\ifx \showDOI      \undefined \def \showDOI       #1{#1}\fi
\ifx \showISBNx    \undefined \def \showISBNx     #1{\unskip}     \fi
\ifx \showISBNxiii \undefined \def \showISBNxiii  #1{\unskip}     \fi
\ifx \showISSN     \undefined \def \showISSN      #1{\unskip}     \fi
\ifx \showLCCN     \undefined \def \showLCCN      #1{\unskip}     \fi
\ifx \shownote     \undefined \def \shownote      #1{#1}          \fi
\ifx \showarticletitle \undefined \def \showarticletitle #1{#1}   \fi
\ifx \showURL      \undefined \def \showURL       {\relax}        \fi
\providecommand\bibfield[2]{#2}
\providecommand\bibinfo[2]{#2}
\providecommand\natexlab[1]{#1}
\providecommand\showeprint[2][]{arXiv:#2}

\bibitem[me:(2021a)]%
        {me:eth2}
 \bibinfo{year}{Retrieved June, 2021}\natexlab{a}.
\newblock \bibinfo{title}{Ethereum project}.
\newblock
\newblock
\urldef\tempurl%
\url{https://www.ethereum.org/}
\showURL{%
\tempurl}


\bibitem[me:(2021b)]%
        {me:lightning}
 \bibinfo{year}{Retrieved June, 2021}\natexlab{b}.
\newblock \bibinfo{title}{Lightnning network, Scalable, Instant
  Bitcoin/Blockchain Transactions}.
\newblock
\newblock
\urldef\tempurl%
\url{https://lightning.network}
\showURL{%
\tempurl}


\bibitem[Cheng et~al\mbox{.}(2018)]%
        {DBLP:journals/corr/abs-1804-05141}
\bibfield{author}{\bibinfo{person}{Raymond Cheng}, \bibinfo{person}{Fan Zhang},
  \bibinfo{person}{Jernej Kos}, \bibinfo{person}{Warren He},
  \bibinfo{person}{Nicholas Hynes}, \bibinfo{person}{Noah~M. Johnson},
  \bibinfo{person}{Ari Juels}, \bibinfo{person}{Andrew Miller}, {and}
  \bibinfo{person}{Dawn Song}.} \bibinfo{year}{2018}\natexlab{}.
\newblock \showarticletitle{Ekiden: {A} Platform for
  Confidentiality-Preserving, Trustworthy, and Performant Smart Contract
  Execution}.
\newblock \bibinfo{journal}{\emph{CoRR}}  \bibinfo{volume}{abs/1804.05141}
  (\bibinfo{year}{2018}).
\newblock
\showeprint[arXiv]{1804.05141}
\urldef\tempurl%
\url{http://arxiv.org/abs/1804.05141}
\showURL{%
\tempurl}


\bibitem[Dziembowski et~al\mbox{.}(2018)]%
        {DBLP:conf/ccs/DziembowskiFH18}
\bibfield{author}{\bibinfo{person}{Stefan Dziembowski},
  \bibinfo{person}{Sebastian Faust}, {and} \bibinfo{person}{Kristina
  Host{\'{a}}kov{\'{a}}}.} \bibinfo{year}{2018}\natexlab{}.
\newblock \showarticletitle{General State Channel Networks}. In
  \bibinfo{booktitle}{\emph{Proceedings of the 2018 {ACM} {SIGSAC} Conference
  on Computer and Communications Security, {CCS} 2018, Toronto, ON, Canada,
  October 15-19, 2018}}, \bibfield{editor}{\bibinfo{person}{David Lie},
  \bibinfo{person}{Mohammad Mannan}, \bibinfo{person}{Michael Backes}, {and}
  \bibinfo{person}{XiaoFeng Wang}} (Eds.). \bibinfo{publisher}{{ACM}},
  \bibinfo{pages}{949--966}.
\newblock
\urldef\tempurl%
\url{https://doi.org/10.1145/3243734.3243856}
\showDOI{\tempurl}


\bibitem[Li et~al\mbox{.}(2020)]%
        {DBLP:conf/middleware/LiTCYXX20}
\bibfield{author}{\bibinfo{person}{Kai Li}, \bibinfo{person}{Yuzhe~Richard
  Tang}, \bibinfo{person}{Jiaqi Chen}, \bibinfo{person}{Zhehu Yuan},
  \bibinfo{person}{Cheng Xu}, {and} \bibinfo{person}{Jianliang Xu}.}
  \bibinfo{year}{2020}\natexlab{}.
\newblock \showarticletitle{Cost-Effective Data Feeds to Blockchains via
  Workload-Adaptive Data Replication}. In \bibinfo{booktitle}{\emph{Middleware
  '20: 21st International Middleware Conference, Delft, The Netherlands,
  December 7-11, 2020}}, \bibfield{editor}{\bibinfo{person}{Dilma~Da Silva}
  {and} \bibinfo{person}{R{\"{u}}diger Kapitza}} (Eds.).
  \bibinfo{publisher}{{ACM}}, \bibinfo{pages}{371--385}.
\newblock
\urldef\tempurl%
\url{https://doi.org/10.1145/3423211.3425696}
\showDOI{\tempurl}


\bibitem[Miller et~al\mbox{.}(2017)]%
        {DBLP:journals/corr/MillerBKM17}
\bibfield{author}{\bibinfo{person}{Andrew Miller}, \bibinfo{person}{Iddo
  Bentov}, \bibinfo{person}{Ranjit Kumaresan}, {and} \bibinfo{person}{Patrick
  McCorry}.} \bibinfo{year}{2017}\natexlab{}.
\newblock \showarticletitle{Sprites: Payment Channels that Go Faster than
  Lightning}.
\newblock \bibinfo{journal}{\emph{CoRR}}  \bibinfo{volume}{abs/1702.05812}
  (\bibinfo{year}{2017}).
\newblock
\showeprint[arXiv]{1702.05812}
\urldef\tempurl%
\url{http://arxiv.org/abs/1702.05812}
\showURL{%
\tempurl}


\bibitem[Wang et~al\mbox{.}(2021)]%
        {DBLP:conf/sigsoft/WangZLTCLC21}
\bibfield{author}{\bibinfo{person}{Yibo Wang}, \bibinfo{person}{Qi Zhang},
  \bibinfo{person}{Kai Li}, \bibinfo{person}{Yuzhe Tang},
  \bibinfo{person}{Xiapu Luo}, {and} \bibinfo{person}{Ting Chen}.}
  \bibinfo{year}{2021}\natexlab{}.
\newblock \showarticletitle{iBatch: Saving Ethereum Fees via Secure and
  Cost-Effective Batching of Smart-Contract Invocations}.
\newblock \bibinfo{journal}{\emph{CoRR}}  \bibinfo{volume}{abs/2106.08554}
  (\bibinfo{year}{2021}).
\newblock
\showeprint[arxiv]{2106.08554}
\urldef\tempurl%
\url{https://arxiv.org/abs/2106.08554}
\showURL{%
\tempurl}


\end{thebibliography}


\begin{thebibliography}{1}

\bibitem{bowman:reasoning}
M.~Bowman, S.~K. Debray, and L.~L. Peterson.
\newblock Reasoning about naming systems.
\newblock {\em ACM Trans. Program. Lang. Syst.}, 15(5):795--825, November 1993.

\bibitem{braams:babel}
J.~Braams.
\newblock Babel, a multilingual style-option system for use with latex's
  standard document styles.
\newblock {\em TUGboat}, 12(2):291--301, June 1991.

\bibitem{clark:pct}
M.~Clark.
\newblock Post congress tristesse.
\newblock In {\em TeX90 Conference Proceedings}, pages 84--89. TeX Users Group,
  March 1991.

\bibitem{herlihy:methodology}
M.~Herlihy.
\newblock A methodology for implementing highly concurrent data objects.
\newblock {\em ACM Trans. Program. Lang. Syst.}, 15(5):745--770, November 1993.

\bibitem{Lamport:LaTeX}
L.~Lamport.
\newblock {\em LaTeX User's Guide and Document Reference Manual}.
\newblock Addison-Wesley Publishing Company, Reading, Massachusetts, 1986.

\bibitem{salas:calculus}
S.~Salas and E.~Hille.
\newblock {\em Calculus: One and Several Variable}.
\newblock John Wiley and Sons, New York, 1978.

\end{thebibliography}

\appendix
\section{Cost of mode M3 under workloads of top write/read accounts}
The amortized cost per transaction of M3 depends on the number of NW and NR. In $C_3$, a relatively larger NW and smaller NR makes the cost lower. We select $A_i$ who has a larger NW/NR as write-intensive accounts. We apply OCIA to the workload of  top write-intensive accounts that we select. 

\begin{figure}[!htp]
  \subfloat[ Top $10\%$ write intensive accounts with blocks delay ]{
    \includegraphics[width=0.24\textwidth]{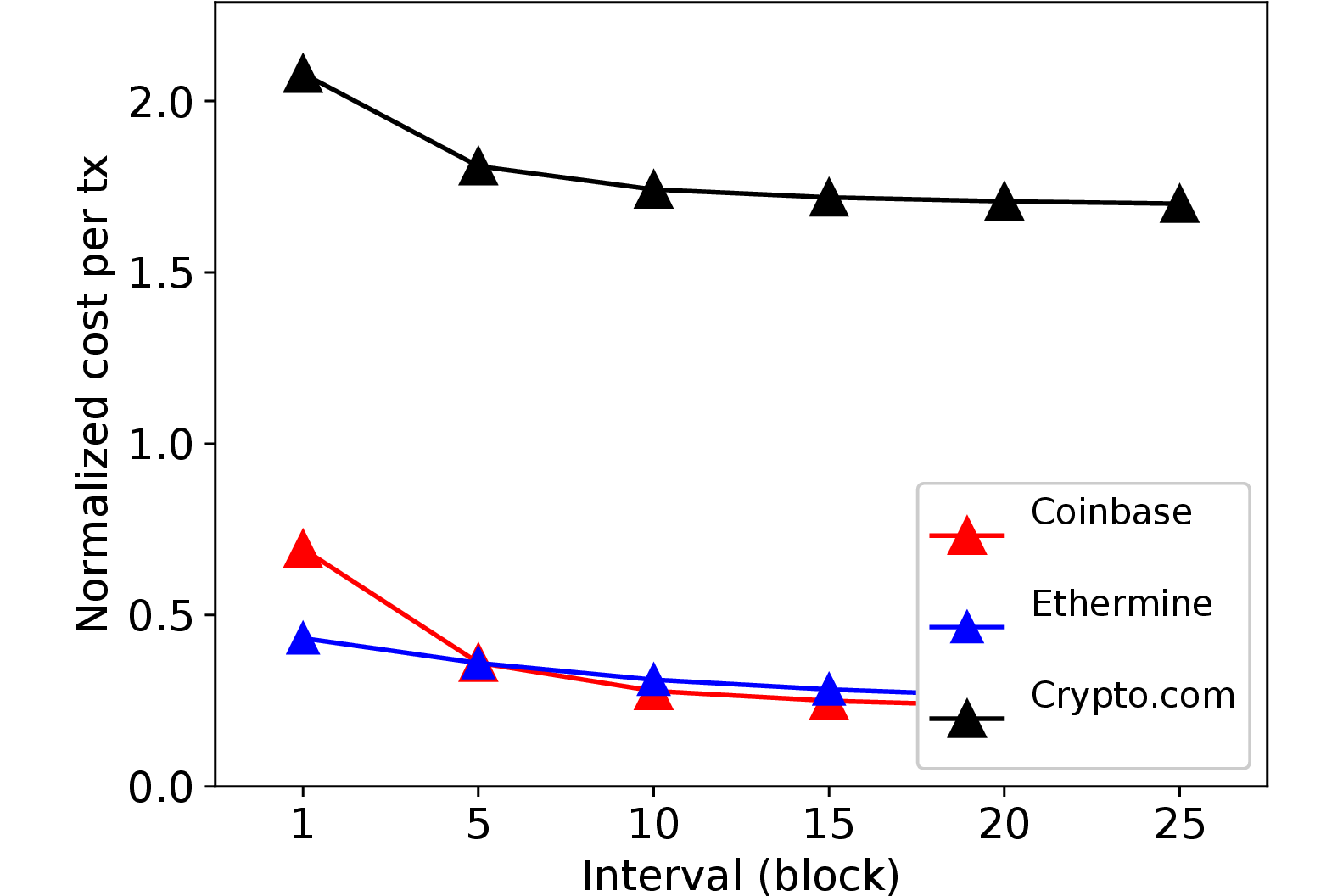}
    \label{fig:top10-cost}
  }
  \subfloat[ Top $x\%$ write-intensive accounts with no delay ]{
    \includegraphics[width=0.24\textwidth]{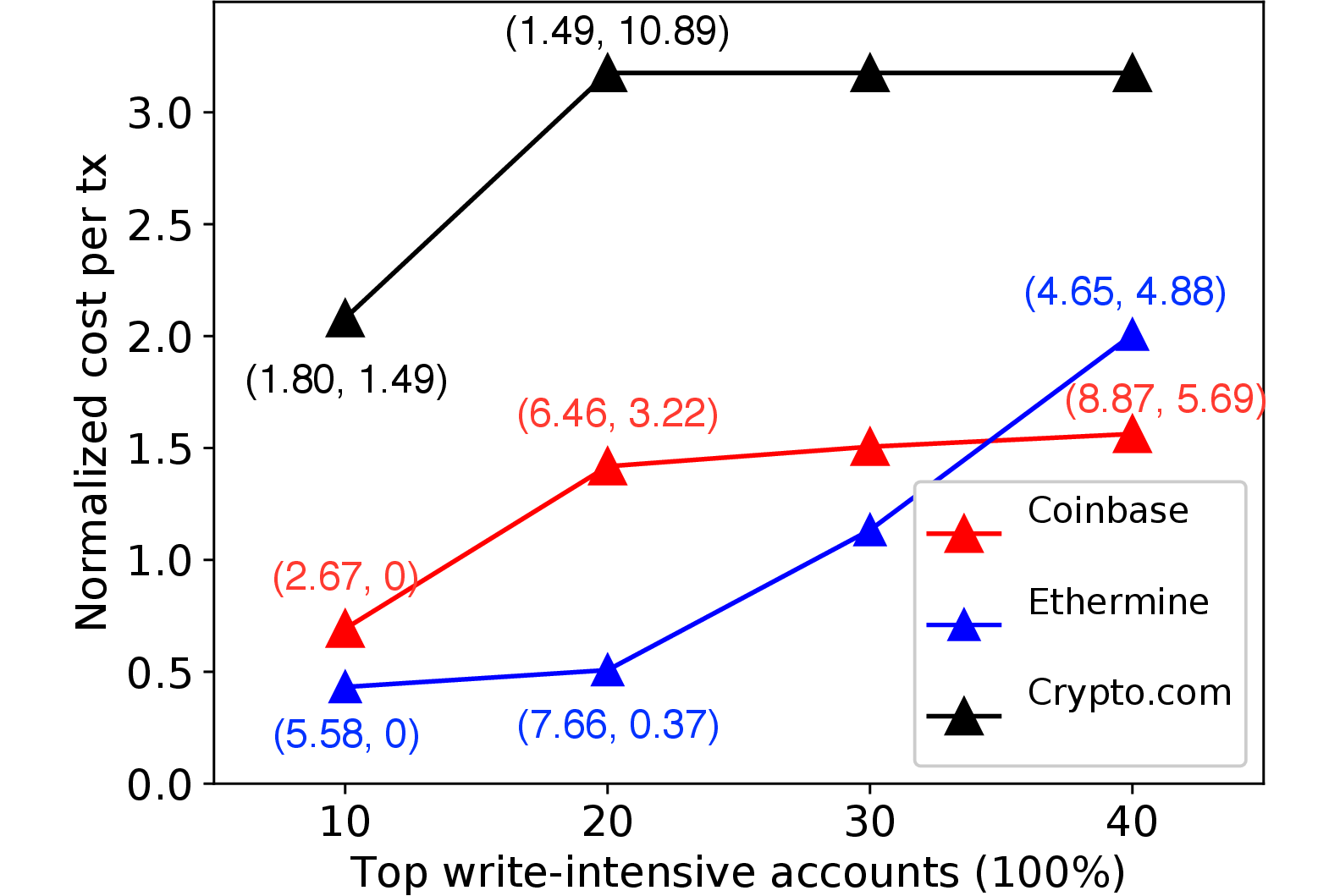}
    \label{fig:topaccounts-nodelay}
  }
  \caption{ The cost of top write intensive accounts workloads}%
\end{figure}

The figure~\ref{fig:top10-cost} illustrates the normalized cost of top $10\%$ write-intensive accounts with various blocks delay. Under the workloads of Coinbase and Ethermine, it saves $30.72\%$ and $56.77\%$ respectively with no delay, while it increases $107.99\%$ overhead under Crypto.com workload. The cost decreases as the block delays increase. It saves $77.34\%$ and $74.87\%$ under the workloads of Coinbase and Crypto.com respectively, while the cost under Crypto.com is still larger than baseline.

The figure~\ref{fig:topaccounts-nodelay} illustrates the normalized cost of top $10\%$, $20\%$, $30\%$ and $40\%$ write-intensive accounts with no delay. We annotate the average of NW and NR of the points in the figure. The result shows only top $10\%$ write-intensive accounts workloads of Coinbase and Ethermine as well as top $20\%$ write-intensive accounts workloads of Ethermine save the cost than baseline, that saves $30.72\%$, $56.77\%$ and $49.23\%$ respectively. 

\end{document}